\documentclass[conference]{IEEEtran}
\IEEEoverridecommandlockouts
\usepackage{cite}
\usepackage[normalem]{ulem}
\usepackage{multirow}
\usepackage{amsmath,amssymb,amsfonts}
\usepackage{algorithmic}
\usepackage{graphicx}
\usepackage{textcomp}
\usepackage{xcolor}
\def\BibTeX{{\rm B\kern-.05em{\sc i\kern-.025em b}\kern-.08em
    T\kern-.1667em\lower.7ex\hbox{E}\kern-.125emX}}
\begin{document}

\title{Bridging the Gap: Integrating Pre-trained Speech Enhancement and Recognition Models for Robust Speech Recognition}

\author{ \IEEEauthorblockN{Kuan-Chen Wang\IEEEauthorrefmark{1}\IEEEauthorrefmark{2}, You-Jin Li\IEEEauthorrefmark{1}\IEEEauthorrefmark{2}, Wei-Lun Chen\IEEEauthorrefmark{2}, Yu-Wen Chen\IEEEauthorrefmark{3}, Yi-Ching Wang\IEEEauthorrefmark{4}, \\ Ping-Cheng Yeh\IEEEauthorrefmark{1}, Chao Zhang\IEEEauthorrefmark{5}, and Yu Tsao\IEEEauthorrefmark{2}}
\IEEEauthorblockA{
\IEEEauthorrefmark{1}National Taiwan University,\
\IEEEauthorrefmark{2}Academia Sinica,\
\IEEEauthorrefmark{3}Columbia University,\
\IEEEauthorrefmark{4}Chunghwa Telecom Co., Ltd.,\
\IEEEauthorrefmark{5}Tsinghua University\\
Email: d12942016@ntu.edu.tw,\
d05942004@ntu.edu.tw,\
renchen8394@gmail.com,\
yu-wen.chen@columbia.edu,\\
ycwang@cht.com.tw,\
pcyeh@ntu.edu.tw,\
cz277@tsinghua.edu.cn,\
yu.tsao@citi.sinica.edu.tw
}
}
\maketitle

\begin{abstract}
Noise robustness is critical when applying automatic speech recognition (ASR) in real-world scenarios. One solution involves using speech enhancement (SE) models as the front end of ASR. However, neural network-based (NN-based) SE often introduces artifacts into the enhanced signals and harms ASR performance, particularly when SE and ASR are independently trained. Therefore, this study introduces a simple yet effective SE post-processing technique to address the gap between various pre-trained SE and ASR models. A bridge module, which is a lightweight NN, is proposed to evaluate the signal-level information of the speech signal. Subsequently, using the signal-level information, the observation adding technique is applied to reduce SE's shortcomings effectively. The experimental results demonstrate the success of our method in integrating diverse pre-trained SE and ASR models, considerably boosting the ASR robustness. Crucially, no prior knowledge of the ASR or speech contents is required during the training or inference stages. Moreover, the effectiveness of this approach extends to different datasets without necessitating the fine-tuning of the bridge module, ensuring efficiency and improved generalization. 

\end{abstract}



\section{Introduction}
Speech enhancement (SE) aims to improve speech quality and intelligibility by recovering clean speech from noisy ones. SE is critical for speech-related applications, such as automatic speech recognition (ASR)~\cite{ochiai2017multichannel,pandey2021dual, sato2022learning,chang2022end} and speaker verification~\cite{michelsanti2017conditional,lee2023lc4sv}, to increase their robustness in real-world environments. Recently, neural network-based (NN-based) methods have become mainstream in SE, owing to the powerful non-linear mapping capabilities of NNs. Various NN-based SE methods have been developed, including multi-layer perceptrons~\cite{lu2013speech,xu2014regression}, convolutional neural networks~\cite{qi2020exploring, 9317794}, fully convolutional networks~\cite{FCN_fu2017raw}, and recurrent neural networks (RNNs)~\cite{RNN_valentini2016investigating,LSTM_weninger2015speech}. Furthermore, numerous studies have proposed advanced NN model architectures or designs, such as DEMUCS~\cite{defossez2020real}, CMGAN~\cite{abdulatif2022cmgan}, MP-SENet~\cite{lu2023mp}, and SEMamba~\cite{chao2024}. NN-based methods exhibit exceptional SE performance compared to conventional methods, demonstrating their potential to enhance downstream speech applications.

ASR is a widely used speech application that converts speech into text. The robustness of ASR is crucial for its practical applicability, where SE can assist as a front end for suppressing noise in input speech~\cite{sato2022learning,chang2022end,9687924,10389733,lee2023d4am}. Despite the impressive performance of NN-based methods, the SE process may introduce artifacts into enhanced signals, potentially harming the ASR performance~\cite{sato2022learning,iwamoto2022bad}. While some studies have attempted the joint training of SE and ASR models to address this issue~\cite{yang2023fat,zhu2023joint}, these methods become infeasible when the cost of training NN models is excessively high or when commercial ASR models are provided in the third part and are inaccessible. Consequently, other studies have focused on the post-processing of SE and have proposed the observation adding (OA) technique~\cite{sato2022learning}. OA involves scaling and combining the original and enhanced speech to form the ASR input. The coefficients of OA in~\cite{sato2022learning} are determined using a switching module. Training the switching module requires adequate training data with the corresponding transcriptions. This process incurs additional costs and may have limited generalizability to other pre-trained SE or ASR models.

To mitigate artifacts while ensuring generalizability, this study introduces a simple yet effective bridge module to determine the OA coefficient. The bridge module is a lightweight NN trained without using information from the backend ASR, and it determines the OA coefficient based on the similarity between the original and enhanced speech. The experimental results demonstrate the effectiveness and generalizability of our method across various SE models, ASR models, and datasets without the need to fine-tune the SE, ASR, and bridge modules. The main contributions of this study are summarized as follows: First, this study proposes an efficient and effective technique for integrating independently pre-trained SE and ASR models. Notably, no fine-tuning is required for either SE or ASR models. Moreover, the bridge module is a lightweight NN model with a simple linear layer, which makes it easy to implement. Second, the proposed method proves effective across different SE models, ASR models, and datasets, and the bridge module requires no fine-tuning.
Third, to the best of our knowledge, this is the first study to apply post-processing to SE to enhance ASR models trained without the need for transcription.

\begin{figure*}[th!]
    \centering
    \includegraphics[width=0.8\textwidth]{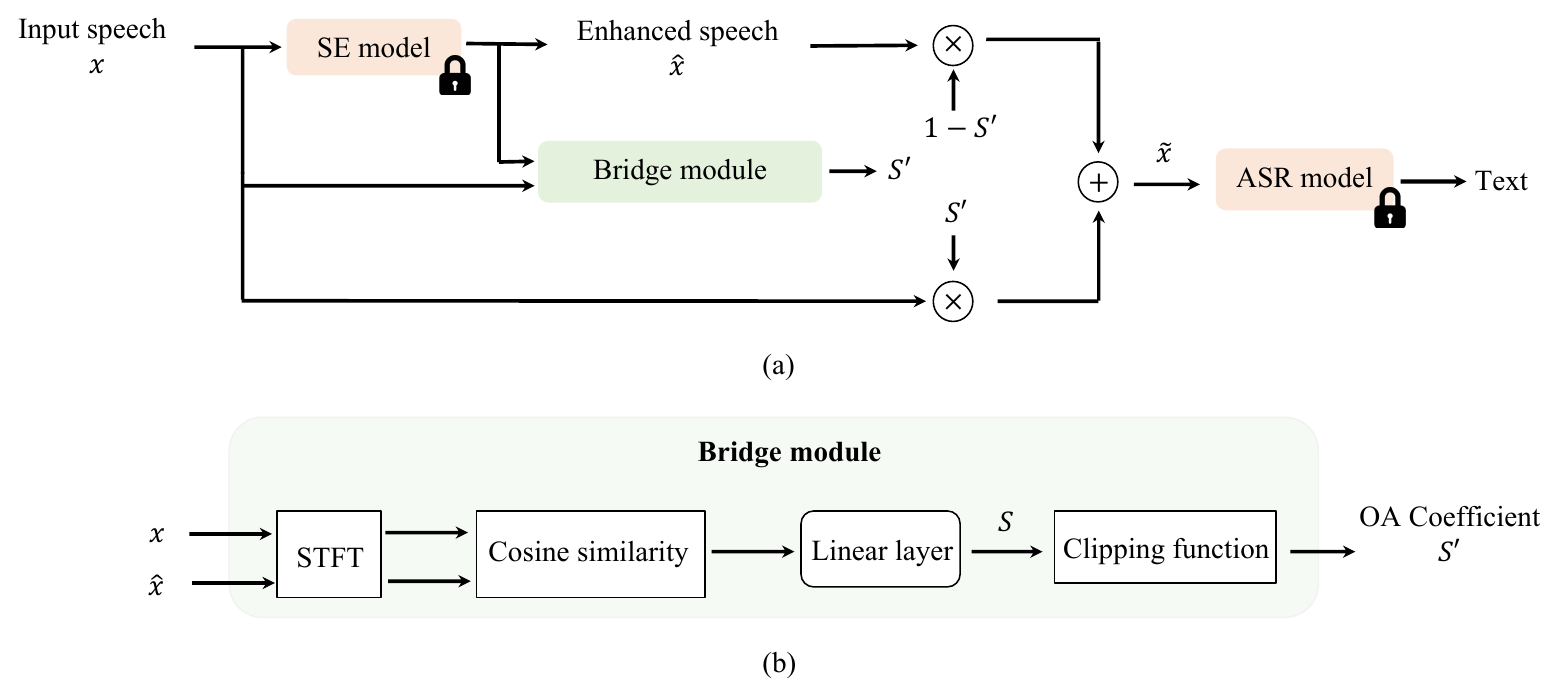}
    \caption{The architecture of the proposed method.}
    \label{fig: Model structure}
\end{figure*}

\section{Related works}
\subsection{SE for noise-robust ASR}
Ensuring the noise robustness of ASR is critical for real-world applications. Previous studies have addressed this challenge by incorporating synthetic or real noisy speech during training to enhance the ASR robustness~\cite{radford2023robust}. For instance, Whisper~\cite{radford2023robust} was trained using a large amount of data collected on the Internet under weak supervision. Despite its inherent noise robustness, ASR can suffer from low signal-to-noise ratios (SNRs). To mitigate potential noise in input speech, previous studies employed SE as a preprocessing stage for ASR~\cite{LSTM_weninger2015speech}. However, NN-based SE methods often introduce artifacts to enhanced signals and potentially deteriorate the ASR performance~\cite{sato2022learning,iwamoto2022bad}. Although the joint training of SE and ASR models has been proposed to improve consistency~\cite{yang2023fat,zhu2023joint}, these methods may be impracticable when commercial ASR models are inaccessible or training costs are unaffordable. Alternatively, some studies have discovered that applying post-processing to SE models, such as OA techniques, can improve ASR performance without previous constraints~\cite{sato2022learning,iwamoto2022bad}. This study also focuses on developing post-processing techniques for robust ASR.

\subsection{Observation adding technique}
OA is a practical technique designed to alleviate speech distortion. In OA, the original and enhanced waveforms are linearly combined to form input signals for downstream tasks. Numerous studies have explored noise-robust speech applications based on OA, including ASR~\cite{sato2022learning}, speaker verification~\cite{lee2023lc4sv}, and speech emotion recognition~\cite{chen2023noise}. The kernel of the OA method lies in determining the coefficient of the waveform combination, which is often achieved through a data-driven process using an NN. Existing NN-based coefficient decision methods, including likelihood-based~\cite{sato2022learning} and reinforcement-learning-based approaches~\cite{lee2023lc4sv}, often leverage information from both the SE and downstream models during training. However, these methods tend to fit specific combinations of SE and downstream models, thereby exhibiting less generalizability. To address this limitation, we propose an OA technique that does not rely on ASR transcription. 

\section{Proposed method}
Fig.~\ref{fig: Model structure} (a) shows the architecture of the proposed robust ASR system. The proposed system comprises three parts: an SE model, an ASR model, and a bridge module. 
\subsection{SE model}
In this study, we employed four well-known pre-trained SE models: CMGAN ~\cite{abdulatif2022cmgan}, MP-SENet~\cite{lu2023mp}, DEMUCS~\cite{defossez2020real}, and SEMamba~\cite{chao2024}. These SE models have diverse properties and can be used to investigate the effectiveness of the proposed method comprehensively. DEMUCS is a waveform-mapping-based SE model~\cite{hao2021denoi,prasad2021investigation}. In contrast, CMGAN, MP-SENet, and SEMamba are spectral-masking-based SE models that achieve state-of-the-art performance on the VoiceBank-DEMAND dataset. Moreover, DEMUCS is trained on the DNS-Challenge dataset~\cite{reddy2020interspeech}. CMGAN, MP-SENet, and SEMamba are trained on the VoiceBank-DEMAND dataset~\cite{botinhao2016speech}. The parameters of these pre-trained SE models remained fixed when applied in this study.

\subsection{ASR model}
Whisper~\cite{radford2023robust} was applied as an ASR model in this study. Whisper is a powerful ASR model trained with a large amount of data under weak supervision and inherently exhibits noise robustness. Various Whisper versions with different model sizes exist. We selected the base and large-v3 versions and fixed their parameters as the SE models in this study.

\subsection{Bridge module}

Fig.~\ref{fig: Model structure} (b) illustrates the structure of the bridge module inspired by a previous study on noise-robust speech emotion recognition~\cite{chen2023noise}. The bridge module first calculates the cosine similarity between the spectrogram of the input and enhanced signals with respect to the time dimension. A linear layer is then used to predict the SNR level~\cite{chen2023noise} $S$. The final output $S'$ used as the OA coefficient is constrained by a predefined clipping function, limiting its value within a specific range (e.g., 0.6 to 1). The design of the bridge module is based on the concept that ASR can perform well with the original speech at high SNR levels, where the enhanced and original spectrograms are similar. Moreover, the floor of the clipping function is specified, as we consider that  Whisper ASR models are inherently noise-robust to some extent. In this study, the floor value was set to 0.6.

The key advantage of the bridge module lies in its generalization capability, as it can be applied to various SE and ASR models without fine-tuning. The training data for the bridge module contained pure clean speech, pure background noise, and enhanced signals, which were generated using CMGAN in this study. The labels for speech and noise signals were set to 1 and 0, respectively. Formulated as a regression problem, the bridge module can learn the distribution of input speech with different noise levels in an unsupervised manner.

\subsection{Integrating pre-trained SE and ASR systems}
The integration of the pre-trained SE, pre-trained ASR, and bridge modules for a robust ASR is presented in this subsection. The SE model transforms the noisy speech waveform $x$ into the enhanced speech waveform $\hat{x}$. The bridge module then predicts the coefficient for OA, $S’$, based on the spectrograms of the noisy speech $x$ and enhanced speech $\hat{x}$. The OA process is subsequently employed to produce the input waveform for ASR, $\tilde{x}$, using the coefficient $S’$. Finally, the ASR model generates the corresponding text. This process can be expressed by the following equations:
\begin{equation}
\label{eq:SE}
    \hat{x}=\text{SE}(x),
\end{equation}
\begin{equation}
\label{eq:Bridge}
    S' =\text{Bridge}(x,\hat{x}),
\end{equation}
\begin{equation}
\label{eq:OA}
    \tilde{x}= S'x+(1-S')\hat{x},
\end{equation}
where SE and Bridge denote the SE model and bridge module, and $x$, $\hat{x}$, and $\tilde{x}$ denote the original, enhanced, and combined speech waveforms, respectively.

\section{Experimental setup}
\subsection{Dataset}
This study adopts clean speech from the LibriSpeech train-360 corpus~\cite{panayotov2015librispeech} and 80\% of the noise data from the DNS-Challenge~\cite{reddy2020interspeech} to train the bridge module. LibriSpeech train-360 contains 103,093 utterances from 921 speakers, and the DNS-Challenge comprises 65,303 foreground and background noise samples.

To evaluate the robustness of the ASR system, noisy utterances are sourced from three datasets: Librispeech with DNS-challenge, Aurora-4~\cite{parihar2004performance}, and the VoiceBank-DEMAND dataset~\cite{botinhao2016speech}. For LibriSpeech, the test-clean corpus (2,580 clean utterances) is adopted, and each utterance is contaminated by five randomly selected noise types from the remaining 20\% of DNS-Challenge at five SNR levels (0, ±6, and ±12 dB). Aurora-4 comprises 3960 noisy utterances, which are from 330 clean utterances distorted by six types of noise and two types of room impulse responses at an SNR in the range of 5-15 dB. The VoiceBank-DEMAND test set comprises 824 noisy utterances at four SNR levels (2.5, 7.5, 12.5, and 17.5 dB).

\subsection{Implementation detail}

The sampling rate of all signals was set to 16 kHz. The STFT parameter setup comprised a Hanning window with a window length of 400 and a hop length of 100. To train the bridge module, we employed a pre-trained CMGAN~\cite{abdulatif2022cmgan} as the SE model. The batch size was 32, and the learning rate was 0.0001. An SGD optimizer with a momentum of 0.9 was adopted. The loss criterion was the mean square error (MSE). 

\subsection{Evaluation criteria}
The performance of ASR is evaluated by the word error rate (WER), where a lower WER denotes better ASR performance. The calculations of two commonly used SE evaluation criteria, perceptual evaluation of speech quality (PESQ)~\cite{PESQ} and short-time objective intelligibility (STOI)~\cite{STOI}, are also provided for reference. Furthermore, we implement a comparison method, SNR-level~\cite{chen2023noise}, following the settings in a previous study with the floor of the clipping function set to zero~\cite{chen2023noise}.

\section{Results and discussion}

\subsection{Predictions of the bridge module}
\begin{figure}[t]
    \centering
    \includegraphics[width=0.7\columnwidth]{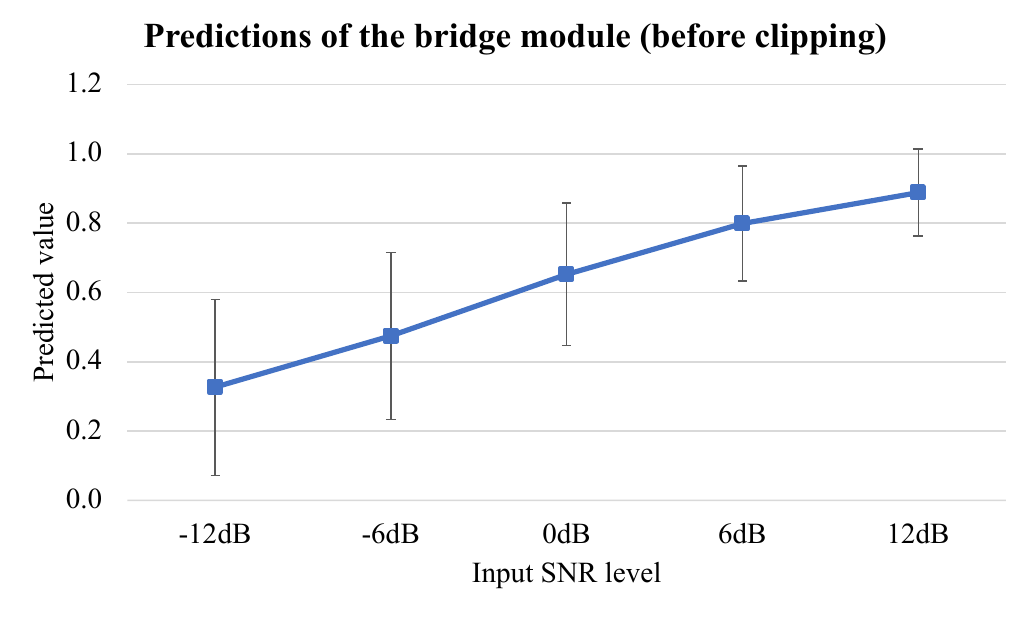}
    \caption{The predictions of the bridge module on the test set, Librispeech with DNS challenge.}
    \label{fig: Predictions}
\end{figure}

Fig.~\ref{fig: Predictions} presents the predictions of the bridge module before clipping on the test dataset of LibriSpeech with DNS-Challenge. We calculate the averages and standard deviations of the predicted values for different SNRs. The predicted values positively correlate with the SNR levels, indicating that the bridge module can effectively model SNR variations in the input signals.

\begin{table*}[ht]
\caption{Evaluation results on LibriSpeech with DNS-challenge test set. Four SE models (CMGAN, MP-SENet, DEMUCS, and SEMamba) are integrated with two Whisper ASR models. (B) and (L) denote the results of Whisper base and large-v3, respectively.}
\centering
\begin{tabular}{cccccccccccccccc}
\hline
\multirow{2}{*}{SE model} & \multirow{2}{*}{Method} & \multicolumn{10}{c}{WER (\%) under different SNRs}                                                                                   & \multicolumn{2}{c}{\multirow{2}{*}{\begin{tabular}[c]{@{}c@{}}Average WER \\(\%)\end{tabular}}} & \multirow{2}{*}{STOI} & \multirow{2}{*}{PESQ} \\ \cline{3-12}
                          &                         & \multicolumn{2}{c}{-12 dB} & \multicolumn{2}{c}{-6 dB} & \multicolumn{2}{c}{0 dB} & \multicolumn{2}{c}{6 dB} & \multicolumn{2}{c}{12 dB} & \multicolumn{2}{c}{}                                                                             &                       &                       \\ \hline
                          &                         & (B)          & (L)        & (B)         & (L)        & (B)        & (L)        & (B)        & (L)        & (B)         & (L)        & (B)              & (L)            & -                     & -                     \\ \hline
                          & Clean                   & -            & -          & -           & -          & -          & -          & -          & -          & -           & -          & 5.8              & 2.8            & -                     & -                     \\
                          & Noisy                   & 73.8         & 44.3       & 46.5        & 18.7       & 21.5       & 6.8        & 11.0       & 3.8        & 7.7         & 3.2        & 32.1             & 15.4           & 0.752                 & 1.272                 \\ \hline
\multirow{3}{*}{CMGAN}    & Enhanced                & 80.3         & 69.1       & 57.4        & 42.0       & 29.4       & 15.5       & 14.0       & 5.9        & 8.5         & 3.7        & 37.9             & 27.2           & \textbf{0.780}        & \textbf{1.642}                 \\
                          & SNR-level~\cite{chen2023noise}               & \textbf{70.7}         & 50.4       & \textbf{43.5}        & 22.7       & \textbf{19.3}       & \textbf{6.8}        & \textbf{10.4}       & \underline{\textbf{3.7}}        & \textbf{7.5}         & \underline{\textbf{3.1}}        & \textbf{30.3}             & 17.4           & \textbf{0.771}        & \textbf{1.302}                 \\
                          & Bridge module           & \textbf{70.3}         & \textbf{43.7}       & \textbf{42.5}        & \textbf{18.0}       & \textbf{19.1}       & \textbf{6.4}        & \textbf{10.4}       & \underline{\textbf{3.7}}        & \textbf{7.5}       & \underline{\textbf{3.1}}        & \textbf{29.9}             & \textbf{15.0}           & \textbf{0.770}        & \textbf{1.298}                 \\ \hline
\multirow{3}{*}{MPSE-Net} & Enhanced                & 91.5         & 81.4       & 74.5        & 62.9       & 49.2       & 35.0       & 25.6       & 14.1       & 13.6        & 6.3        & 50.9             & 39.9           & 0.661                 & \textbf{1.443}                 \\
                          & SNR-level~\cite{chen2023noise}               & 75.6         & 53.7       & 51.3        & 29.1       & 23.5       & 9.3        & \textbf{11.0}       & 3.9        & \textbf{7.5}         & \textbf{3.2}        & 33.8             & 19.8           & \textbf{0.747}        & \textbf{1.309}                 \\
                          & Bridge module           & \textbf{72.3}         & 44.9       & \textbf{45.6}        & 18.9       & \textbf{20.5}       & 6.9        & \textbf{10.7}       & \textbf{3.8}        & \textbf{7.5}        & \textbf{3.2}        & \textbf{31.3}             & 15.5           & \textbf{0.762}        & \textbf{1.305}                 \\ \hline
\multirow{3}{*}{DEMUCS}   & Enhanced                & 77.4         & 61.7       & 51.7        & 32.1       & 25.8       & 11.5       & 13.4       & 5.1        & 8.6         & 3.5        & 35.4             & 22.8           & \underline{\textbf{0.833}}        & \underline{\textbf{1.769}}                 \\
                          & SNR-level~\cite{chen2023noise}               & \textbf{70.3}         & 50.6       & \textbf{40.8}        & 20.1       & \underline{\textbf{18.6}}       & \textbf{6.3}        & \textbf{10.4}      & \underline{\textbf{3.7}}        & \textbf{7.6}         & \underline{\textbf{3.1}}        & \textbf{29.5}             & 16.8           & \textbf{0.788}        & \textbf{1.314}                 \\
                          & Bridge module           & \underline{\textbf{68.4}}        & \underline{\textbf{42.3}}       & \textbf{40.5}        & \underline{\textbf{16.7}}       & \textbf{18.7}       & \underline{\textbf{6.2}}        & \textbf{10.5}       & \underline{\textbf{3.7}}        & \textbf{7.6}         & \underline{\textbf{3.1}}        & \underline{\textbf{29.1}}             & \underline{\textbf{14.4}}           & \textbf{0.777}        & \textbf{1.296}                 \\ \hline
\multirow{3}{*}{SEMamba}   & Enhanced                & 80.4         & 69.5       &    56.9    & 29.6       & 28.6       & 12.9       & 13.2       & 5.2        & 8.2         & 3.3        & 37.5             & 24.1          & \textbf{0.784}       & \textbf{1.760}                 \\
                          & SNR-level~\cite{chen2023noise}        & 77.6       & 65.2       & 52.3       & 36.3       & \textbf{22.6}       & 9.8        & \textbf{10.3}       & 3.9      & \underline{\textbf{7.2}}         & \underline{\textbf{3.1}}        & 34.0             & 23.7           & \textbf{0.781}        & \textbf{1.536}                 \\
                          & Bridge module           & \textbf{70.4}         & \textbf{43.7}       & \textbf{42.1}        & \textbf{17.9}      & \underline{\textbf{18.6}}       & \textbf{6.3}       & \underline{\textbf{10.0}}       & \underline{\textbf{3.7}}        & \textbf{7.3}         & \underline{\textbf{3.1}}        & \textbf{29.7}             & \textbf{15.0}          & \textbf{0.777}        & \textbf{1.370}                 \\ \hline
\multicolumn{16}{l}{\textbf{Bold} indicates that the results outperform noisy baseline, and \underline{underline} denotes the best performance for each specific condition.}                                                                                       
\end{tabular}
\label{tab:unseenSE}
\end{table*}

\subsection{Evaluation results on different SE models}

Table~\ref{tab:unseenSE} summarizes the ASR performance of the different methods with CMGAN, MP-SENet, DEMUCS, and SEMamba on the test set of LibriSpeech with the DNS-challenge. It can be observed that the proposed technique outperforms the other methods under most conditions, proving that SE artifacts can be effectively mitigated by the bridge module. Compared to noisy signals, the proposed method can reduce the overall WERs for the Whisper base by 6.9\% (from 32.1\% to 29.9\%) with CMGAN, 2.5\% (from 32.1\% to 31.3\%) with MP-SENet, 9.3\% (from 32.1\% to 29.1\%) with DEMUCS, and 7.5\% (from 32.1\% to 29.7\%) with SEMamba. The overall WER reductions for Whisper large-v3 are 2.6\% (from 15.4\% to 15.0\%) with CMGAN, 6.5\% (from 15.4\% to 14.4\%) with DEMUCS, and 2.6\% (from 15.4\% to 15.0\%) with SEMamba.

The WER reductions are particularly notable when compared to the enhanced signals. Specifically, CMGAN achieved WER reductions of 21.1\% (from 37.9\% to 29.9\%) for Whisper base and 44.9\% (from 27.2\% to 15.0\%) for Whisper large-v3; MP-SENet achieved reductions of 38.5\% (from 50.9\% to 31.3\%) for Whisper base and 61.2\% (from 39.9\% to 15.5\%) for Whisper large-v3; DEMUCS achieved reductions of 17.8\% (from 35.4\% to 29.1\%) for Whisper base and 36.9\% (from 22.8\% to 14.4\%) for Whisper large-v3; SEMamba achieved reductions of 20.8\% (from 37.5\% to 29.7\%) for Whisper base and 37.8\% (from 24.1\% to 15.0\%) for Whisper large-v3. Furthermore, we observe that the bridge module is more robust than the SNR-level method, confirming the importance of considering the inherent noise robustness of Whisper ASR models. 

Note that speech signals enhanced by CMGAN, SEMamba, MP-SENet, and DEMUCS all obtain higher PESQ and STOI scores than their SNR-level and bridge module counterparts. These results suggest that acquiring the highest PESQ and STOI may not guarantee better recognition results when using Whisper ASR models.

\begin{table}[ht!]
\centering
\caption{Evaluation results on the Aurora-4 test set.}
\label{tab:unseenData}
\begin{tabular}{ccclccc}
\hline
\multirow{2}{*}{SE model} & \multirow{2}{*}{Method}   & \multicolumn{3}{c}{WER (\%)}                                       & \multicolumn{1}{l}{\multirow{2}{*}{STOI}} & \multirow{2}{*}{PESQ} \\ \cline{3-5}
                          &                           & \multicolumn{2}{c}{(B)}                 & (L)                      & \multicolumn{1}{l}{}                      &                       \\ \hline
\multicolumn{1}{l}{}      & Clean                     & \multicolumn{2}{l}{18.6}                & \multicolumn{1}{l}{15.6} & -                                         & -                     \\
                          & Noisy                     & \multicolumn{2}{c}{22.1}                & 17.3                     & 0.812                                     & 1.394                 \\ \hline
\multirow{3}{*}{CMGAN}    & Enhanced                  & \multicolumn{2}{c}{23.6}                & 18.2                     & {\underline{\textbf{0.869}}}                      & \textbf{1.796}        \\
                          & SNR-level~\cite{chen2023noise}                 & \multicolumn{2}{c}{\textbf{21.7}}       & 17.4                     & \textbf{0.846}                            & \textbf{1.627}        \\
                          & Bridge module             & \multicolumn{2}{c}{\textbf{22.0}}       & \textbf{17.2}            & \textbf{0.859}                            & \textbf{1.940}        \\ \hline
\multirow{3}{*}{MP-SENet} & Enhanced                  & \multicolumn{2}{c}{24.4}                & 18.9                     & \textbf{0.864}                            & {\textbf{1.968}}  \\
                          & SNR-level~\cite{chen2023noise}                 & \multicolumn{2}{c}{\textbf{21.8}}       & 17.5                     & \textbf{0.854}                            & \textbf{1.712}        \\
                          & Bridge module             & \multicolumn{2}{c}{{\textbf{21.4}}} & \textbf{17.2}            & \textbf{0.842}                            & \textbf{1.618}        \\ \hline
\multirow{3}{*}{DEMUCS}   & Enhanced                  & \multicolumn{2}{c}{24.8}                & 18.1                     & {\textbf{0.869}}                     & \textbf{1.721}        \\
                          & SNR-level~\cite{chen2023noise}                 & \multicolumn{2}{c}{\textbf{21.9}}       & \textbf{17.3}            & \textbf{0.854}                            & \textbf{1.679}        \\
                          & Bridge module             & \multicolumn{2}{c}{\textbf{21.7}}       & {\underline{\textbf{17.1}}}      & \textbf{0.845}                            & \textbf{1.613}        \\ \hline
\multirow{3}{*}{SEMamba}   & Enhanced                  & \multicolumn{2}{c}{23.6}                & 17.8                     & {\underline{\textbf{0.876}}}                      & \underline{\textbf{2.015}}        \\
                          & SNR-level~\cite{chen2023noise}                 & \multicolumn{2}{c}{\textbf{21.9}}       & 17.5            & \textbf{0.857}                            & \textbf{1.732}        \\
                          & Bridge module             & \multicolumn{2}{c}{\underline{\textbf{21.4}}}       & {\textbf{17.2}}      & \textbf{0.845}                            & \textbf{1.640}        \\ \hline
\multicolumn{7}{l}{\scriptsize \textbf{Bold} indicates that the results outperform noisy baseline, and} \\ 
\multicolumn{7}{l}{\scriptsize \underline{underline} denotes the best performance for each specific condition.}
\end{tabular}
\end{table}

\begin{table}[ht!]
\centering
\caption{Evaluation results on the VoiceBank-DEMAND test set.}
\label{tab:VCTK}
\begin{tabular}{cclllll}
\hline
\multirow{2}{*}{SE model} &
  \multirow{2}{*}{Method} &
  \multicolumn{3}{c}{WER(\%)} &
  \multirow{2}{*}{STOI} &
  \multicolumn{1}{c}{\multirow{2}{*}{PESQ}} \\ \cline{3-5}
                        &               & \multicolumn{2}{c}{(B)}          & \multicolumn{1}{c}{(L)} &                & \multicolumn{1}{c}{} \\ \hline
\multicolumn{1}{l}{}    & Clean         & \multicolumn{2}{l}{6.1}         & \multicolumn{1}{l}{1.8} & -             & -                     \\                        
                        & Noisy         & \multicolumn{2}{l}{9.0}          & 3.1                     & 0.921          & 1.973                \\ \hline
\multirow{3}{*}{CMGAN} & Enhanced & \multicolumn{2}{l}{{\underline{\textbf{7.4}}}} & \textbf{2.8} & \textbf{0.958} & \textbf{3.416} \\
                        & SNR-level~\cite{chen2023noise}     & \multicolumn{2}{l}{\textbf{7.8}} & \textbf{2.8}            & \textbf{0.935} & \textbf{2.196}       \\
                        & Bridge module & \multicolumn{2}{l}{\textbf{7.8}} & \textbf{2.7}            & \textbf{0.933} & \textbf{2.183}       \\ \hline
\multirow{3}{*}{MP-SENet} & Enhanced & \multicolumn{2}{l}{\textbf{8.1}} & \textbf{2.7} & {\textbf{0.960}} & {\textbf{3.499}} \\
                        & SNR-level~\cite{chen2023noise}     & \multicolumn{2}{l}{\textbf{7.8}} & {\underline{\textbf{2.5}}}      & \textbf{0.937} & \textbf{2.263}       \\
                        & Bridge module & \multicolumn{2}{l}{\textbf{8.2}} & \textbf{2.6}            & \textbf{0.934} & \textbf{2.238}       \\ \hline
\multirow{3}{*}{DEMUCS} & Enhanced      & \multicolumn{2}{l}{9.7}          & 3.6                     & \textbf{0.929} & \textbf{2.528}       \\
                        & SNR-level~\cite{chen2023noise}     & \multicolumn{2}{l}{\textbf{8.4}}          & \textbf{2.9}            & \textbf{0.935} & \textbf{2.296}       \\
                        & Bridge module & \multicolumn{2}{l}{\textbf{8.5}} & \textbf{2.9}            & \textbf{0.931} & \textbf{2.223}       \\ \hline
\multirow{3}{*}{SEMamba} & Enhanced      & \multicolumn{2}{l}{\textbf{7.9}}          & \textbf{2.7}                     & \underline{\textbf{0.961}} & \underline{\textbf{3.548}}       \\
                        & SNR-level~\cite{chen2023noise}     & \multicolumn{2}{l}{\underline{\textbf{7.4}}}          & \underline{\textbf{2.5}}            & \textbf{0.942} & \textbf{2.439}       \\
                        & Bridge module & \multicolumn{2}{l}{\textbf{7.6}} & \textbf{2.7}            & \textbf{0.936} & \textbf{2.278}       \\ \hline
\multicolumn{7}{l}{\scriptsize \textbf{Bold} indicates that the results outperform noisy baseline, and} \\ 
\multicolumn{7}{l}{\scriptsize \underline{underline} denotes the best performance for each specific condition.}
\end{tabular}
\end{table}

\subsection{Evaluation results on unseen datasets}

Subsequently, experiments were conducted using two unseen datasets to evaluate the proposed bridge module. Table~\ref{tab:unseenData} lists the results of the Aurora-4 test set in which the proposed method performed best under all conditions. Compared to the use of noisy signals, the bridge module achieved WER reductions for Whisper base of 0.5\% (from 22.1\% to 22.0\%) with CMGAN, 3.2\% (from 22.1\% to 21.4\%) with MP-SENet, 1.8\% (from 22.1\% to 21.7\%) with DEMUCS, and 3.2\% (from 22.1\% to 21.4\%) with SEMamba. For Whisper large-v3, the WER reductions were 0.6\% (from 17.3\% to 17.2\%) with CMGAN, 0.6\% (from 17.3\% to 17.2\%) with MP-SENet, 1.2\% (from 17.3\% to 17.1\%) with DEMUCS, and 0.6\% (from 17.3\% to 17.2\%) with SEMamba. Similar to the results from the Librispeech+DNS Challenge, higher PESQ and STOI scores did not yield better recognition results when using Whisper ASR models.

Table~\ref{tab:VCTK} summarizes the results of the VoiceBank-DEMAND test set. Compared to noisy signals, the proposed method yielded WER reductions of 13.3\% (from 9.0\% to 7.8\%) with CMGAN, 8.9\% (from 9.0\% to 8.2\%) with MP-SENet, 5.6\% (from 9.0\% to 8.5\%) with DEMUCS, and 15.6\% (from 9.0\% to 7.6\%) with SEMamba. For Whisper large-v3, the WER reductions were 12.9\% (from 3.1\% to 2.7\%) with CMGAN, 16.1\% (from 3.1\% to 2.6\%) with MP-SENet, 6.5\% (from 3.1\% to 2.9\%) with DEMUCS, and 12.9\% (from 3.1\% to 2.7\%) with SEMamba. Enhanced speech with higher PESQ and STOI scores cannot guarantee better recognition results.

\section{Conclusion}
In this study, we proposed a post-processing technique that integrates pre-trained SE and ASR models to enhance the robustness of ASR. The bridge module, consisting of a linear layer, determines the OA coefficient based on the similarity between the enhanced and original speech signals. Our experimental results demonstrated that the proposed method considerably improved the noise robustness of ASR when various SE models were applied at the front end. Notably, its effectiveness extended to unseen datasets without fine-tuning any model. In the future, we intend to explore the application of the proposed bridge module to other speech-processing tasks.

\bibliographystyle{IEEEbib}
{
\bibliography{refs}

\begin{thebibliography}{10}

\bibitem{ochiai2017multichannel}
T.~Ochiai, S.~Watanabe, T.~Hori, and J.~R. Hershey,
\newblock ``{Multichannel End-To-End Speech Recognition},''
\newblock in {\em Proc. ICML}, 2017.

\bibitem{pandey2021dual}
A.~Pandey, C.~Liu, Y.~Wang, and Y.~Saraf,
\newblock ``{Dual Application Of Speech Enhancement For Automatic Speech Recognition},''
\newblock in {\em Proc. SLT}, 2021.

\bibitem{sato2022learning}
H.~Sato, T.~Ochiai, M.~Delcroix, K.~Kinoshita, N.~Kamo, and T.~Moriya,
\newblock ``{Learning To Enhance Or Not: Neural Network-based Switching Of Enhanced And Observed Signals For Overlapping Speech Recognition},''
\newblock in {\em Proc. ICASSP}, 2022.

\bibitem{chang2022end}
X.~Chang, T.~Maekaku, Y.~Fujita, and S.~Watanabe,
\newblock ``{End-To-End Integration Of Speech Recognition, Speech Enhancement, And Self-Supervised Learning Representation},''
\newblock {\em arXiv preprint arXiv:2204.00540}, 2022.

\bibitem{michelsanti2017conditional}
D.~Michelsanti and Z.-H. Tan,
\newblock ``{Conditional Generative Adversarial Networks for Speech Enhancement and Noise-Robust Speaker Verification},''
\newblock in {\em Proc. Interspeech}, 2017.

\bibitem{lee2023lc4sv}
C.-C. Lee, H.-W. Chen, C.-S. Chen, H.-M. Wang, T.-T. Liu, and Y.~Tsao,
\newblock ``{LC4SV: A Denoising Framework Learning to Compensate for Unseen Speaker Verification Models},''
\newblock in {\em Proc. ASRU}, 2023.

\bibitem{lu2013speech}
X.~Lu, Y.~Tsao, S.~Matsuda, and C.~Hori,
\newblock ``{Speech Enhancement Based On Deep Denoising Autoencoder.},''
\newblock in {\em Proc. Interspeech}, 2013.

\bibitem{xu2014regression}
Y.~Xu, J.~Du, L.-R. Dai, and C.-H. Lee,
\newblock ``{A Regression Approach To Speech Enhancement Based On Deep Neural Networks},''
\newblock {\em IEEE/ACM Transactions on Audio, Speech, and Language Processing}, vol. 23, no. 1, pp. 7--19, 2014.

\bibitem{qi2020exploring}
J.~Qi, H.~Hu, Y.~Wang, C.-H.~H. Yang, S.~M. Siniscalchi, and C.-H. Lee,
\newblock ``{Exploring Deep Hybrid Tensor-To-Vector Network Architectures For Regression Based Speech Enhancement},''
\newblock {\em arXiv preprint arXiv:2007.13024}, 2020.

\bibitem{9317794}
S.~M. Siniscalchi,
\newblock ``{Vector-to-Vector Regression via Distributional Loss for Speech Enhancement},''
\newblock {\em IEEE Signal Processing Letters}, vol. 28, pp. 254--258, 2021.

\bibitem{FCN_fu2017raw}
S.-W. Fu, Y.~Tsao, X.~Lu, and H.~Kawai,
\newblock ``{Raw Waveform-Based Speech Enhancement By Fully Convolutional Networks},''
\newblock in {\em Proc. APSIPA}, 2017.

\bibitem{RNN_valentini2016investigating}
C.~Valentini-Botinhao, X.~Wang, S.~Takaki, and J.~Yamagishi,
\newblock ``{Investigating RNN-Based Speech Enhancement Methods For Noise-Robust Text-To-Speech.},''
\newblock in {\em Proc. SSW}, 2016.

\bibitem{LSTM_weninger2015speech}
F.~Weninger, H.~Erdogan, S.~Watanabe, E.~Vincent, J.~Le~Roux, J.~R. Hershey, and B.~Schuller,
\newblock ``{Speech Enhancement With LSTM Recurrent Neural Networks And Its Application To Noise-Robust ASR},''
\newblock in {\em Proc. LVA/ICA}, 2015.

\bibitem{defossez2020real}
A.~Defossez, G.~Synnaeve, and Y.~Adi,
\newblock ``{Real Time Speech Enhancement In The Waveform Domain},''
\newblock {\em arXiv preprint arXiv:2006.12847}, 2020.

\bibitem{abdulatif2022cmgan}
S.~Abdulatif, R.~Cao, and B.~Yang,
\newblock ``{CMGAN: Conformer-based Metric-GAN for Monaural Speech Enhancement},''
\newblock {\em IEEE/ACM Transactions on Audio, Speech, and Language Processing}, vol. 32, pp. 2477--2493, 2024.

\bibitem{lu2023mp}
Y.-X. Lu, Y.~Ai, and Z.-H. Ling,
\newblock ``{MP-Senet: A Speech Enhancement Model With Parallel Denoising Of Magnitude And Phase Spectra},''
\newblock {\em arXiv preprint arXiv:2305.13686}, 2023.

\bibitem{chao2024}
R.~Chao, W.-H. Cheng, M.~La~Quatra, S.~M. Siniscalchi, C.-H.~H. Yang, S.-W. Fu, and Y.~Tsao,
\newblock ``{An Investigation of Incorporating Mamba for Speech Enhancement},''
\newblock {\em arXiv preprint arXiv:2405.06573}, 2024.

\bibitem{9687924}
F.-A. Chao, S.-W.~Fan Jiang, B.-C. Yan, J.-w. Hung, and B.~Chen,
\newblock ``{TENET: A Time-Reversal Enhancement Network For Noise-Robust ASR},''
\newblock in {\em Proc. ASRU}, 2021.

\bibitem{10389733}
W.~Zhang, K.~Saijo, Z.-Q. Wang, S.~Watanabe, and Y.~Qian,
\newblock ``{Toward Universal Speech Enhancement For Diverse Input Conditions},''
\newblock in {\em Proc. ASRU}, 2023.

\bibitem{lee2023d4am}
C.-C. Lee, Y.~Tsao, H.-M. Wang, and C.-S. Chen,
\newblock ``{D4AM: A General Denoising Framework for Downstream Acoustic Models},''
\newblock in {\em Proc. ICLR}, 2023.

\bibitem{iwamoto2022bad}
K.~Iwamoto, T.~Ochiai, M.~Delcroix, R.~Ikeshita, H.~Sato, S.~Araki, and S.~Katagiri,
\newblock ``{How Bad Are Artifacts?: Analyzing the Impact of Speech Enhancement Errors on ASR},''
\newblock {\em arXiv preprint arXiv:2201.06685}, 2022.

\bibitem{yang2023fat}
D.~Yang, W.~Wang, and Y.~Qian,
\newblock ``{FAT-HuBERT: Front-End Adaptive Training of Hidden-Unit BERT For Distortion-Invariant Robust Speech Recognition},''
\newblock in {\em Proc. ASRU}, 2023.

\bibitem{zhu2023joint}
Q.-S. Zhu, J.~Zhang, Z.-Q. Zhang, and L.-R. Dai,
\newblock ``{A Joint Speech Enhancement and Self-Supervised Representation Learning Framework for Noise-Robust Speech Recognition},''
\newblock {\em IEEE/ACM Transactions on Audio, Speech, and Language Processing}, vol. 31, pp. 1927--1939, 2023.

\bibitem{radford2023robust}
A.~Radford, J.~W. Kim, T.~Xu, G.~Brockman, C.~McLeavey, and I.~Sutskever,
\newblock ``{Robust Speech Recognition Via Large-scale Weak Supervision},''
\newblock in {\em Proc. ICML}, 2023.

\bibitem{chen2023noise}
Y.-W. Chen, J.~Hirschberg, and Y.~Tsao,
\newblock ``{Noise Robust Speech Emotion Recognition With Signal-to-noise Ratio Adapting Speech Enhancement},''
\newblock {\em arXiv preprint arXiv:2309.01164}, 2023.

\bibitem{hao2021denoi}
Y.~Hao, X.~Huang, H.~Huang, and Q.~Wu,
\newblock ``{Denoi-Spex+: A Speaker Extraction Network Based Speech Dialogue System},''
\newblock in {\em Proc. ICEBE}, 2021.

\bibitem{prasad2021investigation}
A.~Prasad, P.~Jyothi, and R.~Velmurugan,
\newblock ``{An Investigation of End-To-End Models for Robust Speech Recognition},''
\newblock in {\em Proc. ICASSP}, 2021.

\bibitem{reddy2020interspeech}
C.~K.~A. Reddy, V.~Gopal, R.~Cutler, E.~Beyrami, R.~Cheng, et~al.,
\newblock ``{The Interspeech 2020 Deep Noise Suppression Challenge: Datasets, Subjective Testing Framework, And Challenge Results},''
\newblock {\em arXiv preprint arXiv:2005.13981}, 2020.

\bibitem{botinhao2016speech}
C.~V. Botinhao, X.~Wang, S.~Takaki, and J.~Yamagishi,
\newblock ``{Speech Enhancement For A Noise-Robust Text-To-Speech Synthesis System Using Deep Recurrent Neural Networks},''
\newblock in {\em Proc. Interspeech}, 2016.

\bibitem{panayotov2015librispeech}
V.~Panayotov, G.~Chen, D.~Povey, and S.~Khudanpur,
\newblock ``{Librispeech: An Asr Corpus Based On Public Domain Audio Books},''
\newblock in {\em Proc. ICASSP}, 2015.

\bibitem{parihar2004performance}
N.~Parihar, J.~Picone, D.~Pearce, and H.~Hirsch,
\newblock ``{Performance Analysis Of The Aurora Large Vocabulary Baseline System},''
\newblock in {\em Proc. EUSIPCO}, 2004.

\bibitem{PESQ}
A.~W. Rix, J.~G. Beerends, M.~P. Hollier, and A.~P. Hekstra,
\newblock ``{Perceptual Evaluation Of Speech Quality (PESQ)-A New Method For Speech Quality Assessment Of Telephone Networks And Codecs},''
\newblock in {\em Proc. ICASSP}, 2001.

\bibitem{STOI}
C.~H. Taal, R.~C. Hendriks, Richard. Heusdens, and J.~Jensen,
\newblock ``{An Algorithm For Intelligibility Prediction Of Time--frequency Weighted Noisy Speech},''
\newblock {\em IEEE Transactions on Audio, Speech, and Language Processing}, vol. 19, no. 7, pp. 2125--2136, 2011.

\end{thebibliography}
}

\end{document}